\documentclass[12pt]{article}
\usepackage{amssymb}
\usepackage{amsfonts,amssymb,amsmath,amsthm}
\usepackage{epsfig}
\usepackage{slashed}
\usepackage{graphicx}
\usepackage{mathrsfs}
\usepackage{bbm}
\usepackage{color}
\def\and{{\rm and}}

\def\a{\alpha}
\def\b{\beta}

\def\d{\delta}

\def\om{\omega}
\def\p{\partial}
\def\th{\theta}

\def\O{\Omega}

\begin{document}
\renewcommand{\thefootnote}{\fnsymbol{footnote}}
\begin{titlepage}

\vspace{10mm}
\begin{center}
{\Large\bf The coordinate coherent states approach revisited}
\vspace{16mm}

{\large Yan-Gang Miao\footnote{\em E-mail: miaoyg@nankai.edu.cn} and
Shao-Jun Zhang\footnote{\em E-mail: sjzhang@mail.nankai.edu.cn}

\vspace{6mm}
{\normalsize \em School of Physics, Nankai University, Tianjin 300071, \\
People's Republic of China}}

\end{center}

\vspace{10mm}
\centerline{{\bf{Abstract}}}
\vspace{6mm}
We revisit the coordinate coherent states approach through two different quantization procedures in the quantum field theory
on the noncommutative Minkowski plane.
The first procedure, which is based on the normal commutation relation between an annihilation and creation operators,
deduces that a point mass can be described by a Gaussian function
instead of the usual Dirac delta function. However, we argue this
specific quantization by adopting the canonical one (based on the
canonical commutation relation between a field and its conjugate
momentum) and show that a point mass should still be described by
the Dirac delta function, which implies that the concept of point
particles is still valid when we deal with the noncommutativity by
following the coordinate coherent states approach. In order to
investigate the dependence on quantization procedures, we apply the
two quantization procedures to the Unruh effect and Hawking
radiation and find that they give rise to significantly different
results. Under the first quantization procedure, the Unruh
temperature and Unruh spectrum are not deformed by noncommutativity,
but the Hawking temperature is deformed by noncommutativity while
the radiation specturm is untack. However, under the second
quantization procedure, the Unruh temperature  and Hawking
temperature are untack but the both spectra are modified by an
effective greybody (deformed) factor.



\end{titlepage}

\newpage
\renewcommand{\thefootnote}{\arabic{footnote}}
\setcounter{footnote}{0}
\setcounter{page}{2}
\pagenumbering{arabic}

\section{Introduction}
After the seminal work by Seiberg and Witten~\cite{Seiberg1999}, the noncommutativity of spacetimes  was revived
and paid more and more attention to henceforth, for instance, it was believed~\cite{Doplicher1994,Doplicher2003,Garay1994}
to be an indispensable ingredient
for the quantization of gravity.
In general there exist two common methods to deal with the noncommutative spacetime. The first one is the ``star-product",
which encodes the noncommutativity through replacing the ordinary product between functions by the Moyal-Weyl
product~\cite{Weyl1931,Greonewold1946,Moyal1949}.
This method has intensively been applied to the construction of field theories and gravity theories on noncommutative spacetimes. For reviews,
see refs.~\cite{Douglas2001,Szabo2001,Szabo2006,Banerjee2009a}. Within this framework one calculates only order
by order in noncommutative parameters and
then loses the nonlocality of noncommutative theories.
Nevertheless, the second method, i.e. the ``coordinate coherent states approach"~\cite{Smailagic2003a,Smailagic2003b,Smailagic2004} is
quite different from the first one in the study of noncommutative quantum mechanics and quantum field theory.
The models established in quantum field
theory with such an approach are consistent with the Lorentz invariance (restricted in the Euclidean space), unitarity and UV-finiteness.
The main idea of this approach
is that the physical position of a point is represented by the mean value of its coordinate operators on coherent states.
Associated with the quantization procedure based on the normal commutation relation between an annihilation and creation operators,
the plane wavefunction which usually represents a ``free point particle" gets deformed by a damping factor, and then
the Feynman propagator of scalar fields acquires an extra damping factor.
This indicates that a point mass can be described by a Gaussian function
instead of the usual Dirac delta function, i.e. a point mass is smeared over the width $\sqrt{\th}$, where $\th$ is the
noncommutative parameter. Black hole solutions with this Gaussian point source, known as ``noncommutative inspired black holes",
have been shown to possess some special
properties~\cite{Nicolini2005,Rizzo2006,Ansoldi2006,Sapllucci2008,Nicolini2008,Smailagic2010,Modesto2010,Mann2011,Mureika2011,Nicolini2011}.

However, the specific quantization procedure mentioned above is noncanonical, under which,
instead of the commutator between a field and its conjugate momentum,
the commutator between an annihilation and creation operators is imposed to take the normal form as a basic point.
This gives rise to the
result that the commutator between a field and its conjugate momentum takes a deformed form containing a damping factor.
From the point of view of the canonical quantization of field theory, we argue in this paper the noncanonical
quantization procedure by still following the coordinate coherent states approach
but taking the canonical quantization procedure, and quite interestingly we have a significantly different result. That is,
if we demand the commutator between the field and its conjugate momentum takes the canonical
form as the starting point, the Feynman propagator will be the usual form as that in the commutative theory. This indicates
that a point mass should still be described  by the usual Dirac delta function. As a result, the concept of point particles is still valid
if we deal with the spacetime noncommutativity by making use of the coordinate coherent states approach.
We note that the commutator between an annihilation and creation operators now gets deformed in the canonical quantization procedure,
which still leads to deformed modes as the representation of the spacetime noncommutativity.

In order to investigate the dependence on quantization procedures, we apply the two quantization procedures to the Unruh effect
and Hawking radiation and find that they indeed lead to significantly different results. There have been some
literature~\cite{Nicolini2009,Rinaldi2010,Banerjee2008a,Banerjee2008b,Banerjee2009b,Nozari2008,Miao2010} concerning the issues
based on the coordinate coherent states approach associated with the noncanonical quantization procedure. 
For instance, in ref.~\cite{Nicolini2009} the Unruh effect on noncommutative
spacetimes was studied by the introduction of the Unruh-DeWitt detector. It was obtained that
the positive Wightman-Green function acquires an extra damping factor in momentum space and then the response rate is suppressed by
the noncommutativity of spacetime. As a result, it was argued that the Unruh-DeWitt detector registers a temperature which
is so greatly suppressed by the noncommutativity that it can be neglected. In addition,
based on the quantum field theory in curved spacetime
and the Bogoliubov transformation, it was claimed in ref.~\cite{Rinaldi2010} that the Hawking radiation spectrum is not deformed by the noncommutativity
and the Hawking temperature cannot be neglected. We note that the different results caused by the specific treatments in
refs.~\cite{Nicolini2009,Rinaldi2010} emerge from the noncommutativity of spacetime and such a difference disappears on the ordinary (commutative)
spacetime. As to the reasons behind, see ref.~\cite{Rinaldi2010} for the details. We re-examine the issues by using
the coordinate coherent states approach associated with both the noncanonical and canonical
quantization procedures and by considering the technique
of quantum field theory in curved spacetime and the Bogoliubov transformation. The results are shown to be dependent
on the quantization procedures. Under the noncanonical quantization procedure,
the Unruh temperature and Unruh spectrum are not deformed by noncommutativity, but
the Hawking temperature is deformed by noncommutativity while the radiation spectrum is untack. However, under the canonical quantization procedure,
the Unruh  temperature and Hawking temperature are untack but the both spectra are modified
by an effective greybody (deformed) factor.

This paper is organized as follows. In the next section we shall discuss the quantum field theory on the noncommutative Minkowski plane
by means of the coordinate coherent states approach associated with the two quantization procedures.
This section contains two subsections. In the first subsection, we shall give a detailed review on the noncanonical quantization procedure proposed in
refs.~\cite{Smailagic2003a,Smailagic2003b,Smailagic2004,Rinaldi2010}. In the second subsection, we shall take a look at
the coordinate coherent states approach associated with the canonical quantization procedure.
In section 3 we shall discuss the Unruh effect and Hawking radiation by following the two quantization procedures respectively in subsections
3.1 and 3.2.
The final section is devoted to the conclusion.

\section{Quantum field theory on noncommutative Minkowski plane}
In this section, we discuss the quantum field theory on the noncommutative Minkowski plane in terms of the coordinate coherent states approach
under the two quantization procedures. In the noncanonical quantization, the commutator between the annihilation and creation operators
takes the normal
form as the starting point and then the commutator between the field and its conjugate momentum is deformed by noncommutativity.
In the canonical quantization, on the contrary, the commutator between the field and its conjugate momentum is set to be
normal at the beginning and this leads to the noncommutative deformed commutator between the annihilation and creation operators.
Due to the fact that the modes of fields are related closely to the annihilation and creation operators, the noncommutative effects are thus represented
by the deformed modes. From the point of view of quantization of field theory, the latter procedure seems to be more reasonable.
We note that the two quantization procedures are identical if no noncommutativity is involved in.

\subsection{The noncanonical quantization procedure}
In this subsection we just review the treatment and its resulting outcomes under the noncanonical quantization procedure proposed
in refs.~\cite{Smailagic2003a,Smailagic2003b,Smailagic2004,Rinaldi2010}.
For simplicity, we consider a two-dimensional
noncommutative Minkowski plane with coordinate operators $\hat{x}^0$ and $\hat{x}^1$ satisfying the algebra
\begin{equation}
[\hat{x}^0, \hat{x}^1] = i \th, \label{Ncalgebra}
\end{equation}
where $\th$ denotes a constant noncommutative parameter. In terms of complex coordinate operators
$\hat{z} \equiv \hat{x}^0 + i \hat{x}^1$, $\hat{z}^\dagger \equiv \hat{x}^0 - i \hat{x}^1$, the algebra eq.~(\ref{Ncalgebra}) becomes
\begin{equation}
[\hat{z}, \hat{z}^\dagger] = 2 \th.
\end{equation}
The coherent states $|z\rangle$, which are defined as the eigenstates of the complex coordinate operator $\hat{z}$: $\hat{z} |z\rangle = z |z\rangle$,
have the following normalized form,
\begin{equation}
|z\rangle = e^{-\frac{z z^\ast}{4\th}} e^{\frac{z}{2\th} \hat{z}^\dagger}|0\rangle.\label{coherentstate}
\end{equation}
The physical position of a particle $(x^0, x^1)$ is defined as the expectation values of the coordinate operators $(\hat{x}^0, \hat{x}^1)$
on the coherent states,
\begin{equation}
x^0 \equiv \langle z| \hat{x}^0 |z\rangle = {\rm Re} z,\qquad x^1 \equiv \langle z| \hat{x}^1 |z\rangle = {\rm Im} z.\label{xev}
\end{equation}

For the noncommutative version of the plane wave operator $e^{i p_\mu \hat{x}^\mu}$, where $\mu=0, 1$, and the metric is
$\eta_{{\mu}{\nu}}={\rm diag}(-1,1)$, its mean value on the coherent states takes the form\footnote{It can be verified that the damping
factor $e^{-\frac{\th}{4} (p_0^2 + p_1^2)}$ takes the same form in both the Euclidean and Minkowski signatures. Although the commutation
relation between coordinate operators (eq.~(\ref{Ncalgebra}))
preserves the Lorentz boost (rotational) symmetry in the Minkowski (Euclidean) signature,
the noncommutative plane wavefunction eq.~(\ref{planewave}) violates the Lorentz boost invariance.
This violation stems from the Lorentz non-invariance of the coherent states (see eq.~(\ref{coherentstate})).
We point out that such a Lorentz non-invariance is intrinsic in the
coordinate coherent states approach~\cite{Smailagic2003a,Smailagic2003b,Smailagic2004}.
It is easy to check that the coherent states eq.~(\ref{coherentstate}) preserve the Lorentz rotational symmetry
in the Euclidean signature but violate the Lorentz boost symmetry in the Minkowski signature.
For the details, see the Appendix. The reason that we take the Minkowski signature is that we shall discuss the radiation of black holes
by using the technique of quantum field theory in curved spacetime. Therefore, it is natural to choose the Minkowski signature for investigating
relative physical phenomena.},
\begin{equation}
\langle z| e^{ip_0 \hat{x}^0 + i p_1 \hat{x}^1} |z\rangle = e^{-\frac{\th}{4} (p_0^2 + p_1^2) + i p_\mu x^\mu}\label{planewave},
\end{equation}
which is interpreted as the wavefunction of a ``free point particle" on the noncommutative plane. Note that the plane wavefunction acquires a damping
factor $e^{-\frac{\th}{4} (p_0^2 + p_1^2)}$ which will play a central role in quantization.

Now we consider a massless real scalar field $\phi$ propagating on the noncommutative Minkowski plane.
In terms of the coordinate coherent states approach~\cite{Smailagic2003a,Smailagic2003b,Smailagic2004}
the action is undeformed by the spacetime noncommutativity and thus takes the usual form,
\begin{equation}
S= - \frac{1}{2} \int d^2x \ \eta^{\mu\nu} \partial_\mu \phi \partial_\nu \phi,
\end{equation}
which is conformally invariant. The field $\phi$ obeys the Klein-Gordon equation,
\begin{equation}
\p_\mu \p^\mu \phi = 0.
\end{equation}
When quantizing the field, we have to expand the field $\phi({t},{x})$
in terms of a set of plane wavefunction modes. According to the above prescriptions,
on the noncommutative plane the positive frequency modes are modified as
\begin{equation}
u_p (t,x) = \frac{e^{-\th \om^2 /2}}{\sqrt{4\pi\om}} e^{-i\om t+ i p x},\label{mode}
\end{equation}
where we identify $p_0$ with $\om=|p|$ and set $p_1=p$. These modes are orthogonal but have a deformed Klein-Gordon product,
\begin{equation}
(u_p, u_{p'}) \equiv -i \int dx \ (u_p \p_t u_{p'}^\ast - u_{p'}^\ast \p_t u_p) = e^{-\th \om^2} \d(p-p').
\end{equation}
When expanding the field $\phi(t,x)$, we have to use the deformed modes,
\begin{equation}
\phi(t,x) = \int_{-\infty}^\infty dp \ [\hat{a}_p u_p (t,x) + \hat{a}_p^\dagger u_p^\ast (t,x)],\label{field-expansion}
\end{equation}
where the annihilation and creation operators are imposed to satisfy the normal commutation  relation,
\begin{equation}
[\hat{a}_p, \hat{a}_{p'}^\dagger] = \d (p-p').\label{commutator}
\end{equation}
In this case, the equal-time commutator between the field $\phi$ and its conjugate momentum $\dot{\phi}$ is deformed to be
\begin{equation}
[\phi(t,x), \dot{\phi}(t,x')] = \frac{i}{2\sqrt{\pi\th}} e^{-\frac{(x-x')^2}{4\th}}.\label{canonical-commutator}
\end{equation}

Using eqs.~(\ref{field-expansion}) and (\ref{commutator}) we can derive the Wightman's positive frequency function
\begin{equation}
G^+ (x^\mu, x'^{\mu}) \equiv \langle 0|\phi(x^\mu) \phi(x'^{\mu}) |0\rangle = \int \frac{d p}{4\pi\om} e^{-\th\om^2- i \om (t-t')+i p (x-x')}.
\label{Wightman-function}
\end{equation}
As usual, the Feynman propagator $G_F (x^\mu, x'^\mu)$ is defined as
\begin{equation}
G_F (x^\mu, x'^\mu) \equiv \langle 0|T \phi(x^\mu) \phi(x'^{\mu}) |0\rangle = \Theta(t-t') G^+ (x^\mu, x'^\mu)+\Theta(t'-t) G^+ (x'^\mu, x^\mu),
\label{Feynman-propagator}
\end{equation}
where $\Theta(x)$ is the step function. With eq.~(\ref{Wightman-function}) and a proper choice of integral contour, the Feynman propagator
eq.~(\ref{Feynman-propagator}) can be expressed as
\begin{equation}
G_F (x^\mu, x'^\mu) = i \int \frac{d^2 p}{(2\pi)^2} \frac{e^{-\th (p_0^2+p_1^2)/2 - i p_\mu (x^\mu-x'^\mu)}}{p_0^2-p_1^2+i\epsilon}.
\label{Feynman-propagator1}
\end{equation}
Therefore,  we can read off the Feynman propagator in momentum space,
\begin{equation}
G_F (p_0, p_1) = \frac{i}{p_0^2-p_1^2+i\epsilon} e^{- \th (p_0^2+p_1^2)/2}. \label{Feynman-propagator2}
\end{equation}
It is easy to check that it satisfies the following equation,
\begin{equation}
(\p_t^2-\p_x^2) G_F (x^\mu, x'^\mu) = - \frac{i}{2\pi\th} e^{-\frac{(t-t')^2+(x-x')^2}{2\th}}.\label{gauss}
\end{equation}
Note that the damping factor $\exp\{-\frac{(t-t')^2+(x-x')^2}{2\th}\}$ in eq.~(\ref{gauss}) originates from the damping factor
$e^{-\frac{\th}{4} (p_0^2 + p_1^2)}$ in eq.~(\ref{planewave}), and both are independent of the spacetime signatures. For the Euclidean signature,
we only have to replace the temporal coordinate $t$ in eq.~(\ref{gauss}) by one spatial coordinate. We have already emphasized this point in footnote 1.

For a massive real scalar field, we can obtain the same result as eqs.~(\ref{Feynman-propagator1}), (\ref{Feynman-propagator2}) and (\ref{gauss})
just with the replacement of the suitable mass shell condition~\cite{Smailagic2003a,Smailagic2003b,Smailagic2004,Rinaldi2010}.
Note that a Gaussian function appears on the right-hand side of eq.~(\ref{gauss}) instead of the usual Dirac delta function.
The corresponding result\footnote{Actually, eq.~(\ref{Gaussian}) is obtained based on the spacetime with only space-space noncommutativity
where $\th^{0i}=0$. In this situation,
the exponential factor appeared in eq.~(\ref{gauss})
only involves in the spatial coordinates and the Gaussian distribution eq.~(\ref{Gaussian}) is thus deduced.
See also the previous footnote and the comment under eq.~(\ref{gauss}).}
given in refs.~\cite{Nicolini2005,Rizzo2006,Ansoldi2006,Sapllucci2008,Smailagic2010,Modesto2010,Mann2011,Mureika2011,Nicolini2011} is that
a point with mass $M$ is described by the following Gaussian function rather than the usual Dirac delta function
when one considers the noncommutativity abiding by the coordinate coherent states approach under the noncanonical quantization procedure,
\begin{equation}
\rho_{\th}(r)=\frac{M}{\left(4\pi\th\right)^{\frac{3}{2}}}e^{-\frac{r^2}{4\th}}.
\label{Gaussian}
\end{equation}

\subsection{The canonical quantization procedure}
In this subsection we re-examine the issues mentioned in the previous subsection but
follow the canonical quantization procedure. In other words, if we begin with the equal-time canonical commutator between
$\phi$ and $\dot{\phi}$, that is, if eq.~(\ref{canonical-commutator}) is replaced by
\begin{equation}
[\phi(t,x), \dot{\phi}(t,x')] = i \d(x-x'),
\end{equation}
the commutation relation between the annihilation operator and creation operator eq.~(\ref{commutator}) then changes to be
\begin{equation}
[\hat{a}_p, \hat{a}_{p'}^\dagger] = e^{\th \om^2} \d(p-p').\label{deformed-commutator}
\end{equation}
This keeps the same modes of fields (see eqs.~(\ref{mode}) and (\ref{field-expansion})) as in the noncanonical case.
Making a similar analysis to that done in the previous subsection, we can derive the Feynman propagator in momentum space,
\begin{eqnarray}
G_F (p_0, p_1) = \frac{i}{p_0^2 - p_1^2 +i\epsilon}. \label{Feynman-propagator3}
\end{eqnarray}
It is easy to check that the Feynman propagator now satisfies the usual equation,
\begin{equation}
(\p_t^2-\p_x^2) G_F (x^\mu, x'^\mu) = - i \d^2(x^\mu-x'^\mu),\label{green}
\end{equation}
which is not deformed by the noncommutativity of spacetime. We note that a similar result has been mentioned in the star-product
formalism, which provides a helpful support to our result from a different point of view. That is, in the
star-product formalism the
noncommutativity does not affect free fields although it affects interactions. For
more details, see refs.~\cite{Filk1996a,Douglas2001,Szabo2001}.

The authors of the
literature~\cite{Smailagic2003a,Smailagic2003b,Smailagic2004,Nicolini2005}
deduced from the damped factor in eq.~(\ref{gauss}) that the
distribution of ``point mass" can be described by a Gaussian
function (eq.~(\ref{Gaussian})). However, as we see now, when the
canonical quantization procedure is applied, the damped factor
disappears in the Green's equation (eq.~(\ref{green})). As a result,
we can deduce that the ``point mass" is not smeared over the width
$\sqrt{\th}$ and should still be described by the usual Dirac delta
function. This is the key point of the present paper and it does not
depend on the spacetime signatures. The difference of the two Green
functions in eq.~(\ref{gauss}) and eq.~(\ref{green}) leads to some
significant results when we discuss the Unruh effect and Hawking
radiation of the noncommutative inspired Schwarzschild black hole as
we shall show in section 3. We may give a natural interpretation,
i.e.
one can still keep the concept of point particles
when one takes the average effect of noncommutativity\footnote{The characteristic of the coordinate coherent states approach
is the close relationship with the average value of various operators on coherent states. See, for instance, eqs.~(\ref{xev}) and (\ref{planewave}).},
which seems to be a merit for the canonical procedure.
That is,
the effect of noncommutativity is diluted when the coordinate coherent states approach together with the canonical quantization is adopted.

\section{Unruh effect and Hawking radiation on noncommutative Minkowski plane}
In order to investigate the dependence on quantization procedures,
we now use the coordinate coherent states approach to discuss the Unruh effect and Hawking radiation
on the noncommutative Minkowski plane under the two quantization procedures. As we shall see, the two quantization
procedures will give rise to significantly different results. Under the noncanonical procedure, the Unruh temperature and Unruh spectrum
are not deformed by noncommutativity, but the Hawking temperature is deformed by noncommutativity while the radiation spectrum is untack.
Under the canonical procedure, the Unruh temperature and Hawking temperature are untack but the both spectra are modified
by an effective greybody (deformed) factor.

\subsection{Under the noncanonical quantization procedure}
First we give the general formulaes of the Bogoliubov transformation. The left- ($p<0$) and right-moving ($p>0$) modes do not affect
each other in our following discussions and can be considered separately. For simplicity, in the following we only consider the right-moving mode.
In the quantization of fields on a general spacetime, there exist more than one complete orthonormal set of modes
with which we can expand the field operator $\phi$. Assume that $u_\om (t, x)$ and $v_\O (t, x)$ are two bases of orthonormal
modes of the type\footnote{Henceforth, we use energy $\om$ or $\O$ rather than momentum $p$ as subscripts to label different modes.}
as eq.~(\ref{mode}),
\begin{eqnarray}
(u_\om, u_{\om'})&=& C_\om^2 \d(\om-\om'),\nonumber\\
(v_\O, v_{\O'})&=& C_\O^2 \d(\O-\O'),
\end{eqnarray}
where we use $C_\om$ ($C_\O$) to denote the damping factor $C_\om = e^{-\th \om^2/2}$ ($C_\O = e^{-\th \O^2/2}$). As done in ref.~\cite{Rinaldi2010},
we can rewrite the damped modes as $u_\om = C_\om U_\om$ and $v_\O = C_\O V_\O$, where $U_\om$ and $V_\O$
are the standard modes of the commutative theory.

We expand the scalar field $\phi(t,x)$ as
\begin{equation}
\phi (t, x) = \int_0^\infty d \om \ (\hat{a}_\om u_\om + \hat{a}_\om^\dagger u_\om^\ast) =
\int_0^\infty d \O \ (\hat{b}_\O v_\O + \hat{b}_\O^\dagger v_\O^\ast),\label{field-expansion1}
\end{equation}
where the both sets of operators $\hat{a}_\om$ $(\hat{a}_\om^\dagger)$ and $\hat{b}_\O$ $(\hat{b}_\O^\dagger)$ are imposed to
satisfy the normal commutation relations,
\begin{equation}
[\hat{a}_\om, \hat{a}_{\om'}^\dagger] = \d(\om-\om'),\qquad [\hat{b}_\O, \hat{b}_{\O'}^\dagger]=\d(\O-\O').\label{commutator1}
\end{equation}
There exists the so-called Bogoliubov transformation which connects the two sets of operators,
\begin{equation}
\hat{b}_\O = \frac{1}{C_\O} \int_0^\infty d \om \  C_\om (\a_{\O \om}^\ast \hat{a}_\om - \b_{\O\om}^\ast \hat{a}_\om^\dagger),\label{Bogolubov}
\end{equation}
where $\a_{\O\om}$ and $\b_{\O\om}$ are the standard Bogoliubov coefficients given by
\begin{eqnarray}
\a_{\O\om} &=& (V_\O, U_\om),\nonumber\\
\b_{\O\om} &=& - (V_\O, U_\om^\ast).
\end{eqnarray}
We note that the Bogoliubov transformation in ref.~\cite{Rinaldi2010}
is incorrect (see eq.~(24) of ref.~\cite{Rinaldi2010}) although the final result on the Hawking radiation coincides with ours.
Substituting eq.~(\ref{Bogolubov}) into the second equality of eq.~(\ref{commutator1}),
we get the following normalization condition
\begin{equation}
\int_0^\infty d \om \ C_\om^2 (\a_{\O\om}^\ast \a_{\O'\om}-\b_{\O\om}^\ast \b_{\O'\om}) = C_\O^2 \d(\O-\O').\label{normalization}
\end{equation}
When the field is at the ``$a$-vacuum" $|0_a\rangle$ defined as $\hat{a}_\om |0_a\rangle = 0$, the ``$b$-particle"
number measured in the $\O^{\rm th}$
mode $\hat{N}_\O = \hat{b}_\O^\dagger \hat{b}_\O$ is
\begin{equation}
\langle 0_a| \hat{N}_\O |0_a\rangle = \frac{1}{C_\O^2} \int_0^\infty d \om \ C_\om^2 |\b_{\O\om}|^2.\label{spectrum}
\end{equation}

Let us use the above formulaes to re-examine the Unruh effect on the noncommutative Minkowski plane.
For reviews on the Unruh effect on the commutative spacetime,
see refs.~\cite{Unruh1976,Birrell,Mukhanov}.

In the inertial frame $(t, x)$, the metric takes the usual form,
\begin{equation}
ds^2 = dt^2-dx^2 = du dv,
\end{equation}
where $u$ and $v$ are lightcone coordinates defined as
\begin{equation}
u \equiv t-x,\qquad v \equiv t+x.
\end{equation}
The proper set of modes of the type as eq.~(\ref{mode}) corresponding to the inertial observer is
\begin{equation}
u_\om = \frac{e^{-\th \om^2/2}}{\sqrt{4\pi\om}} e^{-i\om u} = C_\om U_\om.
\end{equation}
In the comoving frame $(\eta, \xi)$ of the Rindler observer with a constant acceleration $a$, the metric has the form
\begin{equation}
ds^2 = e^{2 a \xi} (d\eta^2-d\xi^2) = e^{a (\tilde{v}-\tilde{u})} d\tilde{u} d\tilde{v},
\end{equation}
where the lightcone coordinates of the comoving frame are given by
\begin{equation}
\tilde{u} \equiv \eta-\xi,\qquad \tilde{v} \equiv \eta + \xi.
\end{equation}
The proper set of modes of the type as eq.~(\ref{mode}) corresponding to the Rindler observer is
\begin{equation}
v_\O = \frac{e^{-\th \O^2/2}}{\sqrt{4\pi\O}} e^{-i\O \tilde{u}} = C_\O V_\O.\label{mode1}
\end{equation}
The transformation between the inertial frame $(t, x)$ and the comoving frame $(\eta, \xi)$ is given by
\begin{equation}
u=-\frac{1}{a}e^{-a\tilde{u}},\qquad v=\frac{1}{a} e^{a\tilde{v}}.
\end{equation}

We compute the relation between the standard Bogoliubov coefficients $\a_{ij}$ and $\b_{ij}$ and
obtain the well-known result~\cite{Hawking1974,Hawking1975,Birrell,Mukhanov},
\begin{equation}
|\a_{\O\om}|^2 = e^{\frac{2\pi\O}{a}} |\b_{\O\om}|^2.\label{Bogolubov-coefficient}
\end{equation}
For $\O = \O'$ the normalization condition eq.~(\ref{normalization}) becomes
\begin{equation}
\int_0^\infty d\om \ C_\om^2\,(|\a_{\O\om}|^2 - |\beta_{\O\om}|^2) =
C_\O^2 \, \d(0). \label{normalization1}
\end{equation}
From eqs.~(\ref{spectrum}), (\ref{Bogolubov-coefficient}) and (\ref{normalization1}),
we know that when the field is at the ``Minkowski-vacuum" $|0_M\rangle$
defined by the inertial observer $\hat{a}_\om |0_M\rangle=0$, the number of ``$b$-particles" with energy $\O$ observed by the Rindler observer is
\begin{eqnarray}
\langle\hat{N}_\O\rangle \equiv \langle 0_M|\hat{b}_\O^\dagger \hat{b}_\O |0_M\rangle = \frac{1}{C_\O^2} \int_0^\infty d\om\
C_\om^2 |\b_{\om\O}|^2 
= \frac{\d(0)}{e^{\frac{2\pi\O}{a}}-1},\label{spectrum1}
\end{eqnarray}
where $\d (0)$ appears due to an infinite volume $(\d (0) \sim V)$. Therefore, the density of the number of particles takes the form:
$\langle\hat{n}_\O\rangle \equiv \frac{\langle\hat{N}_\O\rangle}{\d (0)} = \frac{1}{e^{\frac{2\pi\O}{a}}-1}$.
From the spectrum eq.~(\ref{spectrum1}) we can read out the Unruh temperature,
\begin{equation}
T_a = \frac{a}{2\pi}.\label{Unruh-temperature}
\end{equation}
This result shows that both the Unruh spectrum and Unruh temperature are not modified by the spacetime noncommutativity.

We now compare our result with  that of ref.~\cite{Nicolini2009} which was obtained based on the Unruh-DeWitt detector.
In ref.~\cite{Nicolini2009} a suppressed response rate in fact corresponds~\cite{Rinaldi2010} to a suppressed energy flux.
In our case, the Hamiltonian operator for the scalar field is
\begin{eqnarray}
\hat{H} &=& \frac{1}{2} \int d\xi \ \left({\dot{\phi}}^2 + {\phi^{\prime}}^2\right)
= \frac{1}{\pi} \int^\infty_0 d\O \ \O \, e^{-\th \O^2}  \left[\hat{b}_\O \hat{b}^\dagger_\O+\hat{b}^\dagger_\O \hat{b}_\O\right]\nonumber\\
&=& \frac{2}{\pi} \int^\infty_0 d\O \ \O \, e^{-\th \O^2}  \left[\hat{N}_\O+\frac{1}{2} \d (0)\right],\label{Hamiltonian}
\end{eqnarray}
where the dot and prime mean derivatives with respect to the Rindler time $\eta$ and Rindler space $\xi$, respectively.
Note that eqs.~(\ref{field-expansion1}), (\ref{commutator1}) and (\ref{mode1}) have been utilized in the derivation of eq.~(\ref{Hamiltonian}).
It is obvious that the energy density of the detected particles is suppressed and so is its flux through the detector.
Due to the finite density of the number of particles mentioned under eq.~(\ref{spectrum1}), the density of zero point energy is also finite
although a divergent term related to $\d (0)$ appears in eq.~(\ref{Hamiltonian}).
A similar case happens again in eq.~(\ref{Hamiltonian2}) which is associated with the canonical quantization procedure and will be
analyzed in the next subsection.
As a consequence, our results are only agreeable with that of ref.~\cite{Nicolini2009} in the aspect of the response rate
but not in the aspect of the Unruh spectrum and Unruh temperature.

We turn to the Hawking radiation of the noncommutative inspired Schwarzschild black hole.
If the point source of the Gaussian function (eq.~(\ref{Gaussian}))
is adopted and the Einstein equation is not modified by the noncommutativity of spacetime, the noncommutative
inspired Schwarzschild black hole solution
takes the form~\cite{Nicolini2005},
\begin{equation}
ds^2=-\left(1-\frac{2M_\th}{r}\right)dt^2+\left(1-\frac{2M_\th}{r}\right)^{-1}dr^2+r^2d\Omega^2,
\end{equation}
where the parameter $M_{\th}$ with the dimension of mass satisfies the following formula,
\begin{eqnarray}
M_{\th}(r)=\int^r_0\rho_\th(r^{\prime})4\pi{r^{\prime}}^2dr^{\prime}=\frac{2M}{\sqrt \pi} \gamma\left(\frac{3}{2},\frac{r^2}{4\th}\right).
\end{eqnarray}
The gamma function in the above equation is defined to be $\gamma (a,b)= \int_0^b \frac{d t}{t} \ t^a e^{-t}$. To simplify our discussion,
we consider a two-dimensional black hole whose metric is same as the time-radial part of the noncommutative inspired Schwarzschild metric.

In accordance with the discussion for the Unruh effect above, we can deduce the Hawking radiation of the two-dimensional noncommutative
inspired Schwarzschild black hole. At present, the Minkowski vacuum is replaced by the so-called ``Kruskal vacuum"
defined by the inertial observer located
at the event horizon and the Rindler observer is replaced by the static observer at infinity~\cite{Mukhanov}. Making the similar computation to
that done for the Unruh effect, we can easily obtain the spectrum of Hawking radiation,
\begin{equation}
\langle \hat{N}_\O \rangle \equiv \langle 0_K| \hat{b}^\dagger_\O \hat{b}_\O |0_K\rangle = \frac{\d(0)}{e^{\frac{2\pi\O}{k}}-1}.
\end{equation}
Note that the spectrum is not modified by the noncommutativity of spacetime but it is related to a modified temperature
that involves in the noncommutativity,
\begin{equation}
T_{\rm H} = \frac{\kappa}{2\pi} = \frac{1}{8\pi M \left[1 -\frac{2M}{\sqrt{\pi\th}} e^{-\frac{M^2}{\th}}
+ \mathcal{O}(\frac{1}{\sqrt{\th}}e^{-\frac{M^2}{\th}})\right]},\label{HT}
\end{equation}
where in the second equality we have calculated the temperature to order $\mathcal{O} (\frac{1}{\sqrt{\th}}e^{-\frac{M^2}{\th}})$.
Incidentally, this deformed Hawking temperature coincides with the result that was obtained
in terms of the Parikh-Wilczek's tunneling method~\cite{Parikh1999}
in refs.~\cite{Banerjee2008a,Banerjee2008b,Banerjee2009b,Nozari2008,Miao2010} .

\subsection{Under the canonical quantization procedure}
Now we reconsider the Unruh effect and Hawking radiation
following the similar analysis to that in the previous subsection but adopting the canonical quantization procedure.
The Bogoliubov transformation eq.~(\ref{Bogolubov})
is still valid while the normalization condition eq.~(\ref{normalization}) changes to be
\begin{equation}
\int_0^\infty d\om \ (\a_{\O\om}^\ast \a_{\O'\om}-\b_{\O\om}^\ast \b_{\O'\om}) = \d(\O-\O').
\end{equation}
The ``$b$-particle" number measured in the $\O^{\rm th}$ mode $\hat{N}_\O = \hat{b}_\O^\dagger \hat{b}_\O$ at the ``$a$-vacuum" is
\begin{eqnarray}
\langle 0_a| \hat{N}_\O |0_a\rangle = \frac{1}{C_\O^2} \int_0^\infty d\om \  |\b_{\O\om}|^2 
= \frac{\d(0)}{e^{\frac{2\pi \O}{a}}-1}e^{\th \O^2} .\label{spectrum2}
\end{eqnarray}
We note that the Unruh spectrum is modified by an effective greybody factor $\Gamma(\O) \equiv e^{\th \O^2}$ which
is related to the noncommutativity of spacetime but the Unruh temperature is not altered. Moreover,
due to eq.~(\ref{Feynman-propagator3}) or
eq.~(\ref{green}) and the reason given below eq.~(\ref{green}),
we obtain an interesting result that the noncommutative inspired Schwarzschild solution goes back to that of
the ordinary one in the commutative theory. Therefore, the Hawking radiation spectrum is also modified by the effective greybody factor
but the Hawking temperature is not altered.

Now let us take a look at the Hamiltonian operator which can be derived in the same way as that of deriving eq.~(\ref{Hamiltonian}).
The Hamiltonian operator corresponding to the canonical quantization procedure takes the form,
\begin{eqnarray}
\hat{H} = \frac{2}{\pi} \int^\infty_0 d\O \ \O \left[e^{- \th \O^2} \hat{N}_\O+\frac{1}{2} \d (0)\right].\label{Hamiltonian2}
\end{eqnarray}
Substituting eq.~(\ref{spectrum2}) into eq.~(\ref{Hamiltonian2}), we can see that the Hamiltonian is same as
the ordinary (commutative) one, which is consistent with the ordinary Feynman propagator (eq.~(\ref{Feynman-propagator3})).
That is, the response rate will not be deformed by the noncommutativity if we start with the Unruh-DeWitt detector by using
the ordinary propagator.

The difference between the two propagators eq.~(\ref{Feynman-propagator2}) and eq.~(\ref{Feynman-propagator3}) emerges from
the difference between the two Hamiltonians eq.~(\ref{Hamiltonian}) and eq.~(\ref{Hamiltonian2}). What we can
confirm now is that it is just the spacetime noncommutativity  that
gives rise to the difference because such a difference disappears when the noncommutative parameter tends to zero.
In fact, the noncanonical quantization procedure is identical with the canonical one
in the quantum field theory on the ordinary (commutative) spacetime.
Although we argue
the noncanonical quantization procedure by adopting the canonical one in the re-examination of the Unruh effect and Hawking radiation,
to single out the preferable one needs further research in both the theoretical and experimental aspects.
At present we may say that the canonical procedure seems to be more reasonable within the framework of quantization of field theory,
and that it is probably interesting to reveal that
the noncommutativity makes the two quantization procedures produce such a difference.
Incidentally, the spacetime  noncommutativity usually gives rise to the violation of properties that hold
on the ordinary (commutative) spacetime, such as the breaking of self-duality in the noncommutative chiral bosons~\cite{Miao2003}.

\section{Conclusion}

In this paper, we revisit the coordinate coherent states approach in the quantum field theory on the noncommutative Minkowski plane by adopting
two quantization procedures. We argue the noncanonical quantization procedure by proposing the canonical one
and find that the Feynman propagator remains the same form as that in the commutative theory. This indicates that the concept of point particles
is still valid, which is different from the result associated
with the noncanonical quantization procedure. We can give a natural explanation
that the effect of noncommutativity is diluted if the noncommutativity is dealt with by the coordinate coherent states approach
associated with the canonical quantization procedure.
In order to see further differences caused by the different quantization procedures, we take the Unruh effect and Hawking radiation as our examples.
In the following we give a summary of the concrete results.

On the one hand, we re-examine the Unruh effect and Hawking radiation on the noncommutative Minkowski plane following
the coordinate coherent states approach associated with the noncanonical quantization procedure.
Based on the Bogoliubov transformation,
we show that the Unruh spectrum (eq.~(\ref{spectrum1})) and Unruh temperature (eq.~(\ref{Unruh-temperature})) keep unchanged
although the general formulae of "$b$-particle" number measured in the $\O^{\rm th}$ mode
(eq.~(\ref{spectrum})) is deformed by the noncommutativity.
In addition, the Hawking radiation spectrum is not deformed but is related to a deformed temperature which involves in the noncommutativity.

On the other hand, if we adopt the canonical quantization procedure, i.e. we begin with the equal-time canonical commutator between
$\phi$ and $\dot{\phi}$, both the Unruh spectrum and Hawking radiation spectrum are deformed by an effective greybody factor
$\Gamma(\O) = e^{\th \O^2}$ although we do not take into account  for the latter the effect of backscattering of the
noncommutative inspired Schwarzschild black hole. When the radiation energy is low ($\O \ll \frac{1}{\sqrt{\th}}$), the greybody factor can be neglected
and the Unruh spectrum and Hawking radiation spectrum will not be modified by the noncommutative parameter $\theta$.
However, when the radiation energy is high enough ($\O \sim \frac{1}{\sqrt{\th}}$), the greybody factor will play an important role.
It is believed that the noncommutative length $\sqrt{\th}$ is of order of the Planck scale $l_P$, so only when the energy of radiation approaches
the Planck energy $\O_P \sim \frac{1}{l_P}$ can the noncommutative effect be manifest.
The Feynman propagator (eq.~(\ref{Feynman-propagator3}))
remains the usual form and a point mass should still be described by the usual Dirac delta function rather than a Gaussian function,
which is contrary to the statement given in
refs.~\cite{Smailagic2003a,Smailagic2003b,Smailagic2004,Nicolini2005,Rizzo2006,Ansoldi2006,Sapllucci2008,Smailagic2010,Modesto2010,
Mann2011,Mureika2011,Nicolini2011}.

The above results can be summarized concisely by the following table:

\begin{center}
\vspace{0.3cm}
  \begin{tabular}{|c|c|c|}
    \hline\parbox[c][1cm][c]{0cm}{}
      Quantities & Noncanonical quantization  & Canonical quantization  \\
    \hline\parbox[c][1cm][c]{0cm}{}
         Commutators & $[\phi(t,x), \dot{\phi}(t,x')] = \frac{i}{2\sqrt{\pi\th}} e^{-\frac{(x-x')^2}{4\th}}$
         & $[\phi(t,x), \dot{\phi}(t,x')] = i \d(x-x')$\\
         ~~~~~~~~~~~ & $[\hat{a}_p, \hat{a}_{p'}^\dagger] = \d (p-p')$ & $[\hat{a}_p, \hat{a}_{p'}^\dagger] = e^{\th \om^2} \d(p-p')$\\
      \hline\parbox[c][1cm][c]{0cm}{}
      $G_F(p_0, p_1)$ & $\frac{i}{p_0^2-p_1^2+i\epsilon} e^{- \th (p_0^2+p_1^2)/2}$ & $\frac{i}{p_0^2 - p_1^2 +i\epsilon}$\\
      \hline\parbox[c][1cm][c]{0cm}{}
        $\langle \hat{N}_\O \rangle$ & $\frac{\d (0)}{e^{\frac{2\pi\O}{a}}-1}$ & $\frac{\d (0)}{e^{\frac{2\pi \O}{a}}-1}e^{\th \O^2}$ \\
     \hline\parbox[c][1cm][c]{0cm}{}
        $\hat{H}$ & $\frac{2}{\pi} \int^\infty_0 d\O \ \O \, e^{-\th \O^2}  \left[\hat{N}_\O+\frac{1}{2} \d (0)\right]$
        & $\frac{2}{\pi} \int^\infty_0 d\O \ \O \left[e^{- \th \O^2} \hat{N}_\O+\frac{1}{2} \d (0)\right]$\\
    \hline
  \end{tabular}
  \vspace{0.3cm}\\
 \end{center}

Recall that the greybody factor for the ordinary Schwarzschild black hole arises from the backscattering of radiations of the gravitational
field and has the following limiting property,
\begin{eqnarray}
\Gamma(\om)\rightarrow \left\{\begin{array}{l l}
1, & \om \gg \frac{1}{M},\\
\frac{A}{4\pi} \om^2, & \om \ll \frac{1}{M},
\end{array}\right. \label{greybody}
\end{eqnarray}
where $A$ is the horizon area of the black hole. We can see that our effective greybody factor has a quite different behavior compared with the
ordinary greybody factor (eq.~(\ref{greybody})). If we take into account both the effects of noncommutativity and of backscattering,
the total effective greybody factor may be very different from the ordinary one (eq.~(\ref{greybody})).
We shall leave it to our further considerations.

Finally, we make some comments. Under the circumstance that
neither the noncanonical nor the canonical quantization procedure
can be excluded by the present research, our results reveal that the
noncommutativity of spacetime leads to some effects that depend on
the quantization procedures in the coordinate coherent states
approach. In fact, it is already well-known that quantization
schemes are ambiguous on noncommutative spacetimes in the
star-product formalism. It has been found~\cite{Bahns2002a} that in
the perturbative formulation of quantum field theory on the
noncommutative Minkowski space, the Feynman and Dyson perturbation
expansions, which are equivalent in commutative quantum field
theory, lead to different results in noncommutative quantum field
theory. For instance, the unitarity is violated in the former, while
it seems to be preserved in the latter. In addition, the analogy
with these models exists in the twist-deformed version of
noncommutative field
theory~\cite{Dimitrijevic2004a,Chaichian2004a,Chaichian2004b,Aschieri2005a,Balachandran2005a,Balachandran2005b}.
There one has to choose either undeformed or deformed commutation
relations for the field oscillators, and the different choices
correspond to different implications, e.g. the former choice gives
some interesting outcomes, such as the cancellation of the UV/IR mixing, the preservation of the Lorentz invariance, and
the violation of the Pauli exclusion principle. When we compare
our case with that of the star-product formalism, we may say that
the choice of either noncanonical or canonical quantization
procedure is analogous to that of either undeformed or deformed
commutation relations. We can see that the similar ambiguity of
choosing the noncanonical or canonical quantization procedure on
noncommutative spacetimes also exists in the coordinate coherent
states approach, and that the different choices also lead to different implications in the Unruh effect and Hawking radiation
as have been demonstrated in the present paper.
The last comment is that we restrict our
discussion to the two-dimensional case. Since a higher
dimensional case can be reduced to a two-dimensional
one~\cite{Iso2006}, our conclusion is expected to be valid in an
arbitrary dimensional spacetime.

\section*{Acknowledgments}
This work was supported in part by the National Natural
Science Foundation of China under grant No.11175090, and by the Fundamental Research Funds for the Central
Universities under grant
No.65030021.
The authors would like to thank the anonymous referee for the
helpful comments which indeed improved this paper greatly.

\section*{Appendix}

Here we give the proof that the coherent states eq.~(\ref{coherentstate}) preserve the Lorentz rotational symmetry
in the Euclidean signature but violate the Lorentz boost symmetry in the Minkowski signature.

For the Euclidean signature that corresponds to $SO(2)$ group,
the coordinate operators $\hat{x}^0$ and $\hat{x}^1$ transform under the rotation as
\begin{eqnarray}
\left(\begin{array}{ll}\hat{x}'^0\\
\hat{x}'^1\end{array}\right) = \left(\begin{array}{ll}\cos\alpha & \sin\alpha\\
-\sin\alpha & \cos\alpha\end{array}\right) \left(\begin{array}{ll} \hat{x}^0\\
\hat{x}^1\end{array}\right),
\end{eqnarray}
where $\alpha$ is the rotational angle.
The constant noncommutative parameter $\theta$ is invariant under the rotation.
Then we can see eq.~(1) is invariant under the rotation,
\begin{eqnarray}
[\hat{x}'^0, \hat{x}'^1] = (\cos^2\alpha + \sin^2\alpha) [\hat{x}^0,\hat{x}^1] = [\hat{x}^0,\hat{x}^1] = i \theta.
\end{eqnarray}
The complex coordinate operators $\hat{z},\hat{z}^\dagger$ transform under the rotation as
\begin{eqnarray}
\hat{z}'=e^{-i\alpha} \hat{z},\qquad \hat{z}'^\dagger = e^{i \alpha} \hat{z}^\dagger.
\end{eqnarray}
Note that $\hat{z},\hat{z}^\dagger$ are not mixed up under the rotation, so we can define universal states for different coordinate operators:
(i) A universal ground state can be defined for different coordinate operators,
$\hat{z} |0\rangle=0$, thus the ground state is invariant under the rotation;
(ii) The coherent states $|z\rangle$ defined by $\hat{z} |z\rangle = z |z\rangle$ (eq.~(3)) are universal for different coordinate operators
with the eigenvalues $z$ transforming as $z'=e^{-i\alpha} z$. As a result, the coherent states are also invariant under the rotation,
which can be proved directly,
\begin{eqnarray}
|z'\rangle =  e^{-\frac{z' z'^\ast}{4\theta}} e^{\frac{z'}{2\theta} \hat{z}'^\dagger}|0\rangle
=  e^{-\frac{z z^\ast}{4\theta}} e^{\frac{z}{2\theta} \hat{z}^\dagger}|0\rangle = |z\rangle.
\end{eqnarray}

While for the Minkowski signature that corresponds to $SO(1,1)$ group, the coordinate operators $\hat{x}^0$ and $\hat{x}^1$
transform under the Lorentz boost transformation as
\begin{eqnarray}
\left(\begin{array}{ll}\hat{x}'^0\\
\hat{x}'^1\end{array}\right) = \left(\begin{array}{ll}\cosh\alpha & -\sinh\alpha\\
-\sinh\alpha & \cosh\alpha\end{array}\right) \left(\begin{array}{ll} \hat{x}^0\\
\hat{x}^1\end{array}\right),
\end{eqnarray}
where $\alpha$ is the boost parameter. We can see that eq.~(1) is also invariant under the Lorentz boost transformation,
\begin{eqnarray}
[\hat{x}'^0, \hat{x}'^1] = (\cosh^2\alpha - \sinh^2\alpha) [\hat{x}^0,\hat{x}^1] = [\hat{x}^0,\hat{x}^1] = i \theta.
\end{eqnarray}
However, the complex coordinate operators $\hat{z},\hat{z}^\dagger$ now mix up under the Lorentz boost transformation,
\begin{eqnarray}
\hat{z}' = \hat{z} \cosh\alpha  -i \hat{z}^\dagger \sinh\alpha ,\qquad \hat{z}'^\dagger = \hat{z}^\dagger \cosh\alpha  + i \hat{z} \sinh\alpha.
\end{eqnarray}
It is analogous to the Bogoliubov transformation. Therefore, we
cannot define a universal ground state. That is, the ground state
for the coordinate operators $\hat{z}$ defined by $\hat{z} |0\rangle
=0$ is no longer the ground state for the transformed coordinate
operators $\hat{z}'$ anymore, i.e., $\hat{z}' |0\rangle = (\hat{z}
\cosh\alpha  -i \hat{z}^\dagger \sinh\alpha) |0\rangle \neq 0$.
Similarly, we cannot expect that the coherent states eq.~(3) are
invariant under the Lorentz boost transformation. We now verify this
outcome by an apagogical proof, that is, if we assume that the
coherent states $|z\rangle$ (eq.~(3)) are invariant under the
Lorentz boost transformation, i.e., $|z\rangle= |z'\rangle$, we then
have
\begin{eqnarray}
\hat{z}' |z'\rangle =\hat{z}' |z\rangle
=(\hat{z} \cosh\alpha  -i \hat{z}^\dagger \sinh\alpha) |z\rangle = (z \cosh\alpha - i \sinh\alpha\ \hat{z}^\dagger) |z\rangle.
\end{eqnarray}
To preserve the property $\hat{z}'|z\rangle \sim |z\rangle$, the last term in the above equation
should have the form $\hat{z}^\dagger |z\rangle \sim |z\rangle$,
that is to say, $|z\rangle$ are also the eigenstates of $\hat{z}^\dagger$. However, it is impossible because one cannot find common eigenstates
for two operators $\hat{z},\hat{z}^\dagger$ which are noncommutative.

We can also understand the above result from a more intuitive and physical point of view.
It is believed that the minimal length
exists due to the noncommutativity of spacetimes.
The minimal length is invariant under the Lorentz rotational transformation in the Euclidean signature but not invariant under the Lorentz boost
transformation in the Minkowski signature.



\begin{thebibliography}{99}
\bibitem{Seiberg1999}
N. Seiberg and E. Witten, {\it String theory and noncommutative geometry}, JHEP {\bf 09} (1999)
032 [arXiv:hep-th/9908142].

\bibitem{Doplicher1994}
S. Doplicher, K. Fredenhagen and J.E. Roberts, {\it Spacetime quantization induced by classical gravity}, Phys. Lett. {\bf B 331} (1994) 39.

\bibitem{Doplicher2003}
S. Doplicher, K. Fredenhagen and J.E. Roberts, {\it The quantum structure of spacetime at the Planck scale and quantum fields},
Commun. Math. Phys. {\bf 172} (1995) 187 [arXiv:hep-th/0303037].

\bibitem{Garay1994}
L.J. Garay, {\it Quantum gravity and minimum length}, Int. J. Mod. Phys. {\bf A 10} (1995) 145 [arXiv:gr-qc/9403008].

\bibitem{Weyl1931}
H. Weyl, {\it The theory of groups and quantum mechanics} (Dover, New York, 1931).

\bibitem{Greonewold1946}
H.J. Groenewold, {\it On the principles of elementary quantum mechanics,} Physica {\bf 12} (1946) 405.

\bibitem{Moyal1949}
J.E. Moyal, {\it Quantum mechanics as a statistical theory,} Proc. Cambridge Phil. Soc. {\bf 45} (1949) 99.

\bibitem{Douglas2001}
M.R. Douglas and N.A. Nekrasov, {\it Noncommutative field theory}, Rev. Mod. Phys. {\bf 73} (2001) 977 [arXiv:hep-th/0106048].

\bibitem{Szabo2001}
R.J. Szabo, {\it Quantum field theory on noncommutative spaces}, Phys. Rept. {\bf 378} (2003) 207 [arXiv:hep-th/0109162].

\bibitem{Szabo2006}
R.J. Szabo, {\it Symmetry, gravity and noncommutativity}, Class. Quant. Grav. {\bf 23} (2006) R199 [arXiv:hep-th/0606233].

\bibitem{Banerjee2009a}
R. Banerjee, B. Chakraborty, S. Ghosh, P. Mukherjee and S. Samanta, {\it Topics in noncommutative
geometry inspired physics}, Found. Phys. {\bf 39} (2009) 1297 [arXiv:0909.1000[hep-th]].

\bibitem{Smailagic2003a}
A. Smailagic and E. Spallucci, {\it Feynman path integral on the noncommutative plane}, J. Phys. {\bf A 36} (2003) L467 [arXiv:hep-th/0307217].

\bibitem{Smailagic2003b}
A. Smailagic and E. Spallucci, {\it UV divergence-free QFT on noncommutative plane}, J. Phys. {\bf A 36} (2003) L517 [arXiv:hep-th/0308193].

\bibitem{Smailagic2004}
A. Smailagic and E. Spallucci, {\it Lorentz invariance, unitarity and UV-finiteness of QFT on noncommutative spacetime}, J. Phys. {\bf A 37} (2004) 1
(Erratum-ibid. {\bf A 37} (2004) 7169) [arXiv:hep-th/0406174].

\bibitem{Nicolini2005}
P. Nicolini, A. Smailagic and E. Spallucci, {\it Noncommutative geometry inspired
Schwarzschild black hole}, Phys. Lett. {\bf B 632} (2006) 547 [arXiv:gr-qc/0510112].

\bibitem{Rizzo2006}
T.G. Rizzo, {\it Noncommutative inspired black holes in extra dimensions}, JHEP {\bf 09} (2006) 021 [arXiv:hep-ph/0606051].

\bibitem{Ansoldi2006}
S. Ansoldi, P. Nicolini, A. Smailagic and E. Spallucci, {\it Noncommutative geometry inspired
charged black holes}, Phys. Lett. {\bf B 645} (2007) 261 [arXiv:gr-qc/0612035].

\bibitem{Sapllucci2008}
E. Spallucci, A. Smailagic and P. Nicolini, {\it Noncommutative geometry inspired higher-dimensional charged black holes},
Phys. Lett. {\bf B 670} (2009) 449 [arXiv:0801.3519[hep-th]].

\bibitem{Nicolini2008}
P. Nicolini, {\it Noncommutative black holes, the final appeal to quantum gravity: a review,} Int. J. Mod. Phys.{\bf A 24} (2009) 1229
[arXiv:0807.1939[hep-th]].

\bibitem{Smailagic2010}
A. Smailagic and E. Spallucci, {\it ``Kerrr" black hole: the lord of the string}, Phys. Lett. {\bf B 688} (2010) 82 [arXiv:1003.3918[hep-th]].

\bibitem{Modesto2010}
L. Modesto and P. Nicolini, {\it Charged rotating noncommutative black holes}, Phys. Rev. {\bf D 82} (2010) 104035
[arxiv:1005.5605[gr-qc]].

\bibitem{Mann2011}
R.B. Mann and P. Nicolini, {\it Cosmological production of noncommutative black holes}, Phys. Rev.{\bf D 84} (2011) 064014 [arXiv:1102.5096[gr-qc]].

\bibitem{Mureika2011}
J.R. Mureika and P. Nicolini, {\it Aspects of noncommutative (1+1)-dimensional black holes}, Phys. Rev. {\bf D 84} (2011) 044020
[arXiv:1104.4120[gr-qc]].

\bibitem{Nicolini2011}
P. Nicolini, G. Torrieri, {\it The Hawking-Page crossover in noncommutative anti-deSitter space,}  JHEP {\bf 1108} (2011) 097 [arXiv:1105.0188[gr-qc]].

\bibitem{Nicolini2009}
P. Nicolini and M. Rinaldi, {\it A minimal length versus the Unruh effect,} Phys. Lett. {\bf B 695} (2011) 303 [arXiv:0910.2860[hep-th]].

\bibitem{Rinaldi2010}
M. Rinaldi, {\it Particle production and transplanckian problem on the non-commutative plane,} Mod. Phys. Lett. {\bf A 25} (2010) 2805
[arXiv:1003.2408[hep-th]].

\bibitem{Banerjee2008a}
R. Banerjee, B.R. Majhi and S. Samanta, {\it Noncommutative black hole thermodynamics}, Phys. Rev. {\bf D 77} (2008) 124035 [arXiv:0801.3583[hep-th]].

\bibitem{Banerjee2008b}
R. Banerjee, B.R. Majhi and S.K. Modak, {\it Noncommutative Schwarzschild black hole and area law}, Class. Quant. Grav. {\bf 26} (2009) 085010
[arXiv:0802.2176[hep-th]].

\bibitem{Banerjee2009b}
R. Banerjee, S. Gangopadhyay and S.K. Modak, {\it Voros product, noncommutative Schwarzschild black hole and corrected area law},
Phys. Lett. {\bf B 686} (2010) 181 [arXiv:0911.2123[hep-th]].

\bibitem{Nozari2008}
K. Nozari and S.H. Mehdipour, {\it Hawking radiation as quantum tunneling from noncommutative Schwarzschild black hole},
Class. Quant. Grav. {\bf 25} (2008) 175015 [arXiv:0801.4074[gr-qc]].

\bibitem{Miao2010}
Y.-G. Miao, Z. Xue and S.-J. Zhang, {\it Tunneling of massive particles from noncommutative Schwarzschild black hole},
Gen. Relativ. Gravit. {\bf 44} (2012) 555 [arXiv:1012.2426[hep-th]].

\bibitem{Filk1996a}
T. Filk, {\em Divergencies in a field theory on quantum space},
Phys. Lett. {\bf B 376} (1996) 53.

\bibitem{Unruh1976}
W.G. Unruh, {\it Notes on black hole evaporation}, Phys. Rev. {\bf D 14} (1976) 870.

\bibitem{Birrell}
N.D. Birrell and P.C.W. Davies, {\it Quantum fields in curved space} (Cambridge University Press, Cambridge, England, 1982).

\bibitem{Mukhanov}
V.F. Mukhanov and S. Winitzki, {\it Introduction to quantum effects in gravity} (Cambridge University Press, Cambridge, England, 2007).

\bibitem{Hawking1974}
S.W. Hawking, {\it Black hole explosions}, Nature {\bf 248} (1974) 30.

\bibitem{Hawking1975}
S.W. Hawking, {\it Particle creation by black holes}, Commun. Math. Phys. {\bf 43} (1975) 199
[Erratum-ibid. {\bf 46} (1976) 206].

\bibitem{Parikh1999}
M.K. Parikh and F. Wilczek, {\it Hawking radiation as tunneling}, Phys. Rev. Lett. {\bf 85} (2000) 5042
[arXiv:hep-th/9907001].

\bibitem{Miao2003}
Y.-G. Miao, H.J.W. M$\ddot{\rm u}$ller-Kirsten and D.K. Park, {\em Chiral bosons in
noncommutative spacetime}, JHEP {\bf 08} (2003) 038 [arXiv:hep-th/0306034].

\bibitem{Bahns2002a}
D. Bahns, S. Doplicher, K. Fredenhagen and G. Piacitelli, {\em On
the unitarity problem in space/time noncommutative theories}, Phys.
Lett. {\bf B 533} (2002) 178 [arXiv:hep-th/0201222].

\bibitem{Dimitrijevic2004a}
M. Dimitrijevic and J. Wess, {\em Deformed bialgebra of
diffeomorphisms}, arXiv:hep-th/0411224.

\bibitem{Chaichian2004a}
M. Chaichian, P. Kulish, K. Nishijima and A. Tureanu, {\em On a
Lorentz-invariant interpretation of noncommutative space-time and
its implications on noncommutative QFT}, Phys. Lett. {\bf B 604}
(2004) 98 [arXiv:hep-th/0408069].

\bibitem{Chaichian2004b}
M. Chaichian, P. Pre$\check{s}$najder and A. Tureanu, {\em New
concept of relativistic invariance in NC space-time: twisted
Poincar¨¦ symmetry and its implications}, Phys. Rev. Lett. {\bf 94}
(2005) 151602 [arXiv:hep-th/0409096].

\bibitem{Aschieri2005a}
P. Aschieri, C. Blohmann, M. Dimitrijevic, F. Meyer, P. Schupp and
J. Wess, {\em A gravity theory on noncommutative spaces}, Class.
Quant. Grav. {\bf 22} (2005) 3511 [arXiv:hep-th/0504183].

\bibitem{Balachandran2005a}
A.P. Balachandran, G. Mangano, A. Pinzul and S. Vaidya, {\em Spin
and statistics on the Groenewold-Moyal plane: Pauli-forbidden levels
and transitions}, Int. J. Mod. Phys. {\bf A 21} (2006) 3111
[arXiv:hep-th/0508002].

\bibitem{Balachandran2005b}
A.P. Balachandran, A. Pinzul and B. Qureshi, {\em UV-IR mixing in
non-commutative plane}, Phys. Lett. {\bf B 634} (2006) 434
[arXiv:hep-th/0508151].

\bibitem{Iso2006}
S. Iso, H. Umetsu and F. Wilczek, {\it Anomalies, Hawking radiations and regularity in rotating black holes},
Phys. Rev. {\bf D 74} (2006) 044017 [arXiv:hep-th/0606018].

\end{thebibliography}
\end{document}